\documentstyle[11pt,newpasp,twoside,epsf]{article}
\begin{document}
\title{Binary Eclipsing Millisecond Pulsars:  A Decade
of Timing}
\author{David J. Nice}
\affil{Physics Department, Princeton University, Box 708, 
Princeton, NJ 08544 USA}
\author{Zaven Arzoumanian}
\affil{NASA-Goddard Space Flight Center, Mailstop 662.0,
Greenbelt, MD 20771 USA}
\author{Stephen E. Thorsett}
\affil{Department of Astronomy and Astrophysics, University
of California, Santa Cruz, CA 95064 USA}

\begin{abstract}
We present results of long-term timing of eclipsing binaries PSR~B1744$-$24A
and PSR~B1957+20 at Arecibo, the VLA, and Green Bank.  Both pulsars exhibit
irregularities in pulsar rotation and orbital motion.  Increases and decreases
of the orbital period of PSR~B1957+20 are of order $\Delta P_b/P_b\sim
10^{-7}$, varying on a time scale of a few years.  Over a decade of
observations, the orbital period of PSR~B1744$-$24A has only decreased, with
time scale $|P_b/\dot{P_b}|\sim200$\,Myr.  When the effects of orbital motion
are removed from the timing data, long-term trends remain in the pulse phase
residuals, with amplitudes of order 30 and 500\,$\mu$s, respectively, for
B1957+20 and B1744$-$24A.  Such large ``timing noise'' is not seen in other spun-up
pulsars (isolated or binary), leading us to conclude that it is a consequence
of mass flow in the system.  Possible causes include variations in the rotation
of the pulsars and movement of the binary systems along the line of sight (perhaps
due to gravitational interactions with outflowing matter).
\end{abstract}

\section{Introduction}

There are now seven known eclipsing binary pulsars (see Table 1).  In each
system the eclipses are longer than expected for a secondary confined to its
Roche lobe.  From this it is inferred that there is continuous mass flow from
the secondary, constantly replenishing a volume larger than its Roche lobe.
The mass loss may be due to Roche lobe overflow, stellar wind, or a
combination, and in some cases is induced by heating of the secondary by
pulsar irradiation.  Observation of ionized material at eclipse edges,
detected as dispersion measure increases, confirms this picture.

\begin{table}
\caption{Eclipsing Binary Pulsars}
\begin{tabular}{c@{}lrcccc}
\tableline
   &                       & \multicolumn{1}{c}{Pulsar} &  Orbital &                 &  & \\
   &                       & \multicolumn{1}{c}{Period} &   Period & ${{\rm M}_2}^*$ &  Globular  &  \\
\multicolumn{2}{c}{Pulsar} & \multicolumn{1}{c}{(ms)}   &    (hr)  & (M$_{\sun}$)    &  Cluster   &  References \\
\tableline
B&1718$-$19   & 1004.04 &   6.2              &       0.13            &  NGC 6342 &     1 \\
B&1744$-$24A  &   11.56 &   1.8              &       0.10            &  Ter 5    &     2, 3 \\
B&1957+20     &    1.61 &   9.2              &       0.03            &           &     4, 5 \\
J&2051$-$0827 &    4.51 &   2.4              &       0.03            &           &     6, 7 \\
\multicolumn{2}{l}{47 Tuc J} &    2.10 &   2.9              &       0.02            &  47 Tuc   &     8 \\
\multicolumn{2}{l}{47 Tuc O} &    2.64 &   3.3              &       0.03            &  47 Tuc   &     8 \\
\multicolumn{2}{l}{47 Tuc R} &    3.48 &   1.5              &       0.03            &  47 Tuc   &     8 \\
\tableline
\tableline
\multicolumn{7}{l}{$^*$ Assuming M$_1$=1.35M$_{\sun}$, inclination $60\deg$.}\\
\multicolumn{7}{l}{References:  1.  Lyne et al. (1993).  2.  Lyne et al. (1990).}\\
\multicolumn{7}{l}{3.  Nice \& Thorsett (1992).  4.  Fruchter, Stinebring, \& Taylor (1989).} \\
\multicolumn{7}{l}{5.  Arzoumanian, Fruchter, \& Taylor (1994)  6.  Stappers et al. (1996).} \\
\multicolumn{7}{l}{7.  Stappers et al. (1998).  8. Camilo et al. (2000).} \\

\end{tabular}
\end{table}

These systems, or systems similar to them, form a crucial link in the
evolution of low-mass neutron star binaries into isolated pulsars.  Common
characteristics include short orbits (2 to 10 hours) and light secondaries
(0.02 to 0.15\,M$_{\sun}$).  Mass loss rates estimated from the dispersion and
attenuation of radio signals traveling through the ionized wind suggest the
secondaries will not be ablated within a Hubble time, unless there is a large,
undetected neutral component to the wind.  The key question of interest is:
how do these systems evolve?  Can they, in fact, evaporate their secondaries,
or are there similar systems that can?  Can planets form from the mass
outflow?  And how are they related to low mass X-ray binaries, such as
SAX\,J1808.4$-$3658 (van der Klis, this volume)?

\begin{figure}
\plotone{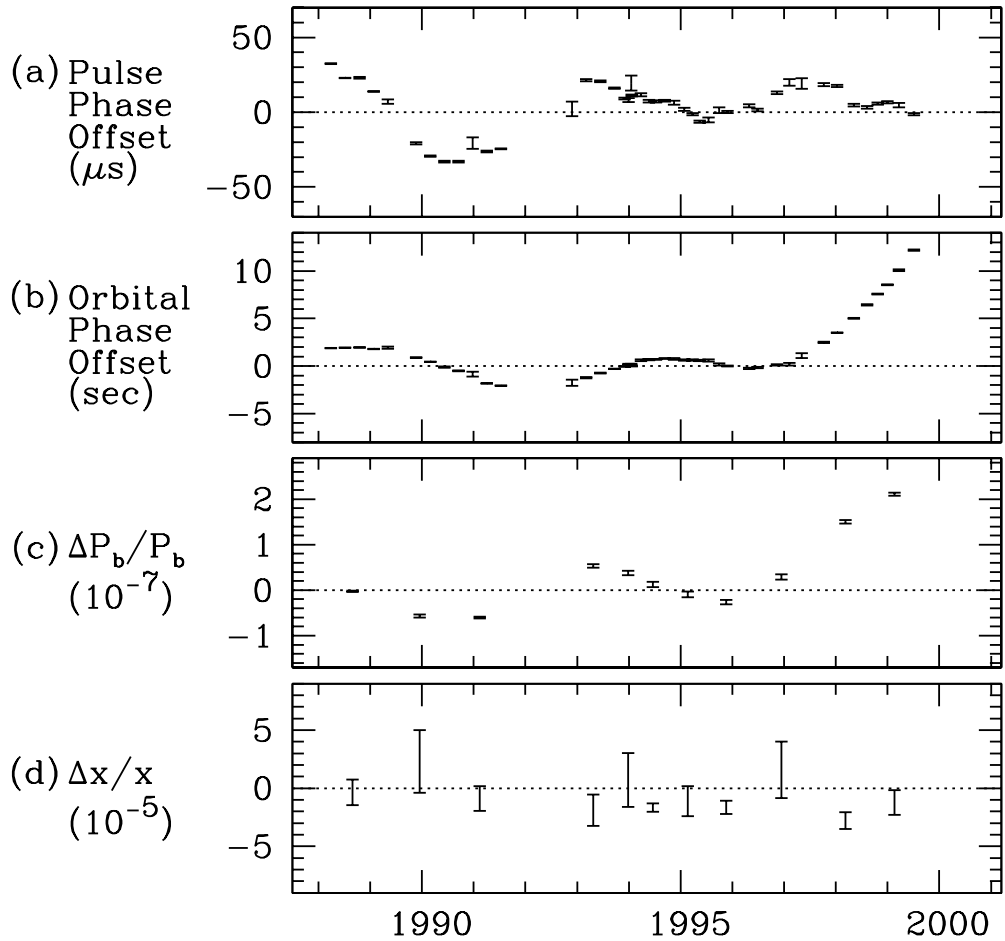}
\caption{Timing results for PSR~B1957+20.  Data through 1994
are Arecibo 430 MHz, after 1994 are Green Bank 575 MHz.
(a,b) Pulse phase offset and orbital phase offset 
at each observing epoch (Green Bank), or averaged over 100 day intervals (Arecibo).
(c,d) Orbital period offset (derivative of figure b) and
orbital size offset, calculated by averaging sets
of four observing epochs (Green Bank) or 400 days (Arecibo).}
\end{figure}

\begin{figure}
\plotone{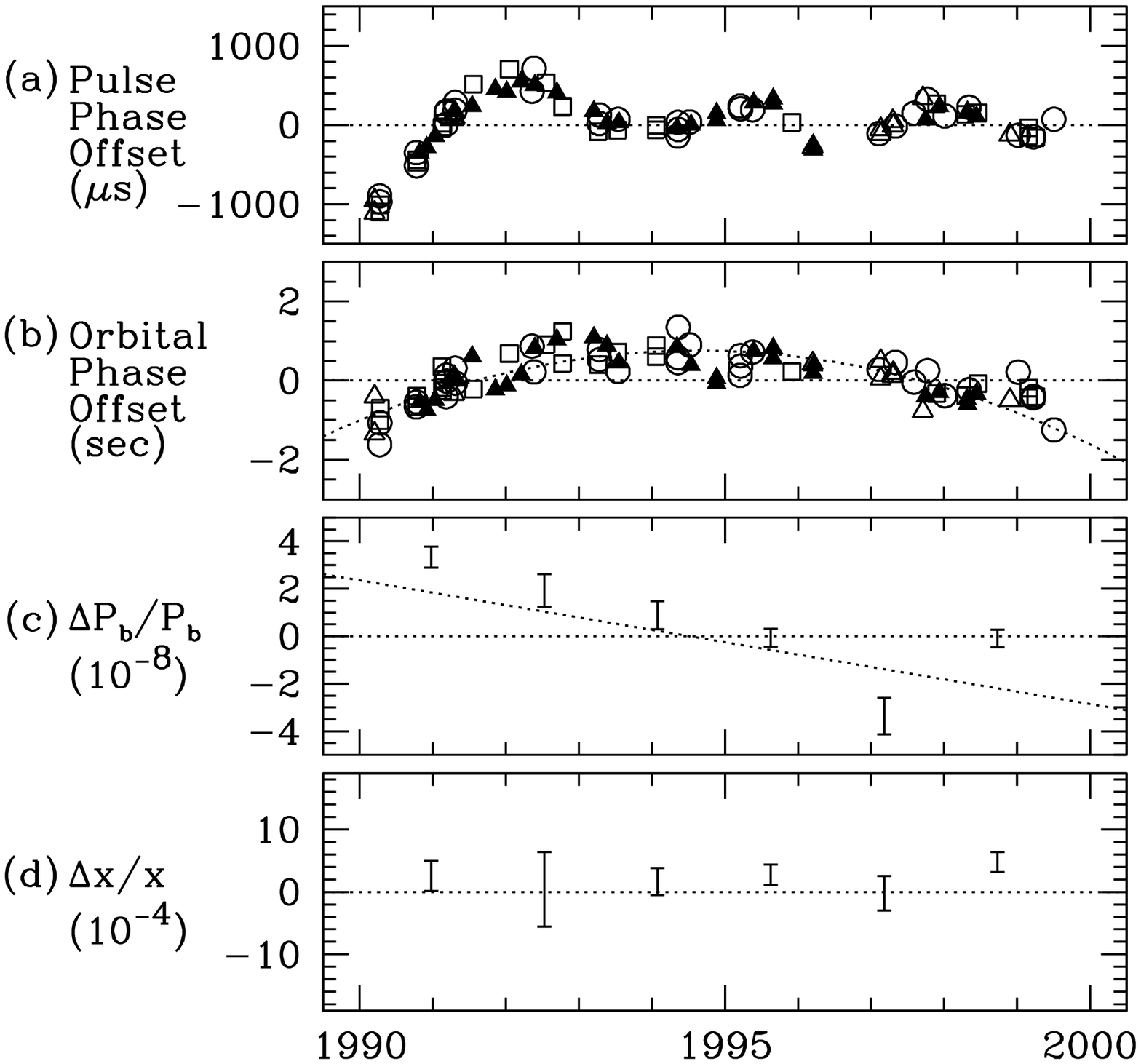}
\caption{Timing results for PSR~B1744$-$24A.  (a,b) Pulse phase
offset and orbital phase offset at each observing epoch.  
Symbols represent Green Bank 800 MHz (circles), Green Bank
1300 MHz (squares) Green Bank 1600 MHz (open triangles), and
VLA 1660 MHz (filled triangles).  Uncertainties are typically 
100\,$\mu$s in pulse phase and 0.6\,s in orbital phase.
(c,d) Orbital period offset (derivative of figure b) and
orbital size offset, calculated by fitting arrival times within
six equal size subsets of data.  Dashed lines show best-fit
model with constant orbital period derivative.}
\end{figure}

\section{Observations}

We have made timing observations of two eclipsing binaries over a span of ten
years.  Observations of PSR~B1957+20 were made at Arecibo at 430\,MHz from
1988 to 1994 and at Green Bank at 575\,MHz from 1994 to 1999.  Observations
of PSR~B1744$-$24A were made at the VLA at 1660\,MHz from 1990 to 1999 and at
Green Bank at 800, 1330, and 1600\,MHz, also from 1990 to 1999.

We used {\sc Tempo} for our initial analysis of pulse arrival times.  After
removing the standard spin-down and Keplerian orbital model, systematic
residual arrival times remained.  We analyzed these residuals by 
fitting small subsets of the data for offsets in pulsar rotation phase,
orbital phase, and orbit size (projected semi-major axis).  For a uniformly
sampled orbit, these phenomena perturb the timing data in orthogonal ways: a
pulse rotation offset uniformly affects arrival times at all orbital phases;
an orbital phase offset induces excess delays proportional to
$\cos(\phi_b)$, where $\phi_b$ is the orbital phase, measured from the
ascending node; and an orbital size change induces excess delay
proportional to $\sin(\phi_b)$.  
At Green Bank and the VLA, entire orbits were observed on a
single day, so we independently analyzed each observing epoch.  Arecibo
observations had poor orbital coverage on any given day, so we analyzed these
data in intervals of 100 days.  Results of this analysis are
given in Figures 1 and 2.

\section{Discussion:  PSR~B1957+20}

The PSR~B1957+20 system is undergoing apparently random variations in orbital period,
with magnitude $\Delta P_b/P_b \sim 10^{-7}$ and a time scale of $\sim
5$\,years (Figure 1c).  This agrees with the conclusion of Arzoumanian,
Fruchter, \& Taylor (1994), who analyzed the Arecibo data from 1988 to 1993.
They found the orbital period first decreased and then increased, and
predicted that the binary period would exhibit ``small quasi-periodic
oscillations.''  A possible source of the period variations is deformation of
the secondary by magnetic activity, which in turn is produced by
dissipation of tidal energy (Applegate \& Shaham 1994).

Modest ``timing noise'' is evident in the pulsar signal (Figure 1a).  This
could be simple variation in dispersion measure---this source has not been
regularly observed at multiple frequencies.  However, similarity to the
PSR~B1744$-$24A timing noise (see below) suggests the origin may lie within the
system itself.

\section{Discussion:  PSR~B1744$-$24A}

The orbital perturbations of PSR~B1744$-$24A, of order $\Delta P_b/P_b \sim
10^{-8}$, are smaller than those of PSR~B1957+20.  A secular downward trend
fits the data reasonably well, though not perfectly (Figures 2b, c).  (The
specific model plotted was derived from a timing fit to
all data, and is essentially equivalent to a quadratic fit to the data in
Figure 2b.)  The evolutionary time scale inferred from this model is
$|P_b/\dot{P_b}|_{\rm obs}=200$\,Myr.  This is within an order of
magnitude of the orbital decay time scale from gravitational
radiation, $|P_b/\dot{P_b}|_{\rm GR}=1000$\,Myr.  It seems likely that
GR plays a substantial role in the evolution of this system.

The pulsar phase residuals of PSR~B1744$-$24A pose an interesting puzzle.
There are systematic variations in pulse arrival time of order $\sim 1$\,ms
(Figure 2a).  These excess residuals are largely independent of frequency.
(Small frequency-dependent differences in arrival times, interpreted as slow
changes in dispersion measure, have been removed from the figure.)  There are
several possible origins of these residuals:

\begin{enumerate}
\item  {\it ``Timing noise'' intrinsic
to the pulsar rotation.}  However, timing noise of
this magnitude is not seen in other spun-up pulsars.  For example,
the relatively noisy residuals of PSR\,B1937+21 reported by Kaspi, Taylor,
\& Ryba (1994) show a peak-to-peak excursion of only $7\,\mu$s over more
than 8 years, more than two orders of magnitude smaller than the residuals
we measure for PSR\,B1744$-$24A.
\item {\it Changes in the viewing geometry of the emission region relative to the
line of sight.}  PSR~B1744$-$24A is expected to precess at a rate of $\sim
1.1\deg$/yr, so a substantial change in viewing angle is possible over 10
years of observations.  However, the pulsar has presumably been spun up by
accretion of mass and angular momentum from the secondary, aligning the pulsar's
angular momentum with the orbital
angular momentum.  With this near alignment, precession effects
would be difficult to detect.  
\item {\it Changes in the distance between the entire binary
system (pulsar and secondary) and the Earth}.  This could
be caused by a lumpy ``excretion disk'', which could serve as a precursor to
planet formation around the pulsar (Banit \& Shaham 1992; Banit et al. 1993).
The data are suggestive towards this end---a periodic 1\,ms signal in
the residuals, with a period of 5\,yr, would correspond to an orbit of a
0.3\,M$_{\earth}$ object about the binary system.  However, there
are no clean periodic signals, so a model like this must
invoke a stochastic ``background'' of lumps (asteroids?).
\item {\it Torques on the pulsar due to infalling matter.}  The
eclipses in this system are highly variable, and the pulsar signal sometimes
disappears for several orbits (or possibly longer).  This suggests mass may be
flowing close to the pulsar, which could be accreted or
interact via the ``propeller effect'', in either case causing a torque on the
pulsar, changing the rotation period.  Such accretion might be
identifiable by X-ray observations of the source at times when the radio
signal is not visible.  Intriguingly, bursts were seen from this general
direction by the Hakucho satellite in 1980 (Makishima et al. 1981), but due to
low angular resolution of this telescope and confusion from sources in the
dense Galactic center region, any connection between the bursts
and PSR B1744$-$24A must be considered speculative.
\end{enumerate}

\acknowledgements
Much of the 1957+20 data from Arecibo was collected by A. S. Fruchter,
M. F. Ryba, and J. H. Taylor.  This work was supported by the National Science
Foundation (United States), via grants to Princeton University and support of
the Arecibo Observatory (operated by Cornell University) and Green Bank and
the VLA (operated by Associated Universities, Inc.).

\end{document}